# Aerosol Jet Printing of High-Temperature Multimodal Sensors for Strain and Temperature Sensing

*Md. Omarsany Bappy, Qiang Jiang, Stephanie Atampugre, and Yanliang Zhang\**

Department of Aerospace and Mechanical Engineering, University of Notre Dame, Notre Dame, Indiana 46556, USA

**KEYWORDS:** multimodal sensors, thermocouples, nanoparticles, aerosol jet printing, printed sensors, high-temperature applications, health monitoring

**ABSTRACT:** Integrating multiple sensing capabilities into a single multimodal sensor greatly enhances its applications for in-situ sensing and structural health monitoring. However, the fabrication of multimodal sensors is complicated and limited by the available materials and existing manufacturing methods that often involve complex and expensive fabrication processes. In this study, a high-temperature multimodal sensor is demonstrated by aerosol jet printing of gold and ITO nanoparticle inks. The printed multimodal sensor for concurrent strain and temperature sensing possesses a high gauge factor of 2.54 and thermopower of 55.64 µV/°C combined with excellent high-temperature thermal stability up to 540 °C. Compared to traditional single-modality sensors, the printed multimodal sensor significantly increases sensing capacity and improves spatial resolution using microscale printed patterns. The study also demonstrates that the strain sensor with integrated thermocouple enables in-situ compensation of the temperature effect on strain sensing, significantly improving strain measurement accuracy at high temperatures. By combining aerosol jet printing with nanomaterial inks, a wide range of multifunctional devices can be developed for a broad range of emerging applications.

## 1. INTRODUCTION

In recent years, there has been significant increasing interest in utilizing sensors across a diverse range of applications such as power generation, manufacturing processes, biometric sensing, and structural health monitoring in industries such as aerospace, nuclear, marine, and civil engineering.[1-5] Increased static or dynamic loads and the widespread adoption of lightweight components necessitate real-time monitoring of strain and temperature on parts with high risks of structural failures.[6, 7] Multifunctional devices with multiple sensing capabilities have attracted considerable interest in structural health monitoring, in situ sensing, human-machine interfaces, soft robots, and wearable sensing applications.[8-14]

Additive manufacturing has emerged as a very attractive method to print sensors for a broad range of applications.[15, 16] A variety of materials and methods have been explored to improve the performance of printed sensors for monitoring physical properties such as strain and temperature. For example, printed thermocouples were made using a variety of metals such as Molybdenum silicide vs. Tungsten silicide[17], Indium oxide vs. Indium Tin oxide[18], and Graphene vs. MXene[19]. Graphite thin film[20, 21], Silver nano particles[22], and MXene ink[19] have been widely used in strain sensing applications. Direct printing techniques such as inkjet, aerosol jet, and screen printing have garnered significant attention in recent years due to their ability to transform nanoscale materials directly into functional devices. Significant progress has been achieved in the realm of printed sensors that possess a single modality, such as strain[20-26], temperature[2, 27-30], and pressure[31-34]. Nevertheless, the fabrication of multimodal sensors remains a challenge owing to the intricate manufacturing processes and the complexity of decoupling multiple signals.[35-39] Aerosol jet printing (AJP) enables direct 3D conformal printing of multimodal sensors onto components with complex geometry, resulting in intimate thermal and mechanical coupling and accurate temperature and strain measurements. A graphene and MXene nano ink-based flexible bimodal sensor by AJP was demonstrated to monitor temperature and detect strain simultaneously up to 150 °C.[19] However, the multimodal sensor for simultaneous measurement of strain and temperature at high temperatures is yet to be explored. Under high-temperature operating conditions, the electrical and mechanical properties of printed sensors can be negatively impacted due to oxidation, fatigue, drifting, thermal deformation, and creep, leading to various challenges.[6, 40, 41]

Herein, we report an aerosol jet printed high-temperature multimodal sensor for simultaneous strain and temperature measurements. The multimodal sensor is fabricated using ITO and Gold nanoparticles, owing to their exceptional thermal stability, oxidation resistance, and consistent sensing performance at high temperatures. The AJP also makes it possible to print sensors with a high spatial resolution (~10 μm).[15, 42-46] The printed sensor studied here shows a gauge factor of 2.54 at room temperature, which is around 30% greater than that of typical metal-based strain gauges (e.g., copper-nickel alloy)[21, 22] with a temperature-sensing thermopower of 55.64 μV/°C.

## 2. EXPERIMENTAL METHODS

**2.1. Sensor Design and Fabrication.** The aerosol jet printing method is utilized to print the multimodal sensor in this study. AJP offers the ability to print sensors with fine features, including line widths as small as ~10 μm and film thicknesses ~100 nm, thereby enabling the integration of multiple sensor materials into a compact device. This printing method also allows non-intrusive sensor implementation, featuring intimate thermal contacts and mechanical coupling, leading to highly precise temperature and strain measurements. Furthermore, AJP has the capability to print sensors on not only 2D surfaces but also irregular-shaped components with curved surfaces, such as valves and welded joints, which are more susceptible to failure and are difficult to attach conventional sensors.

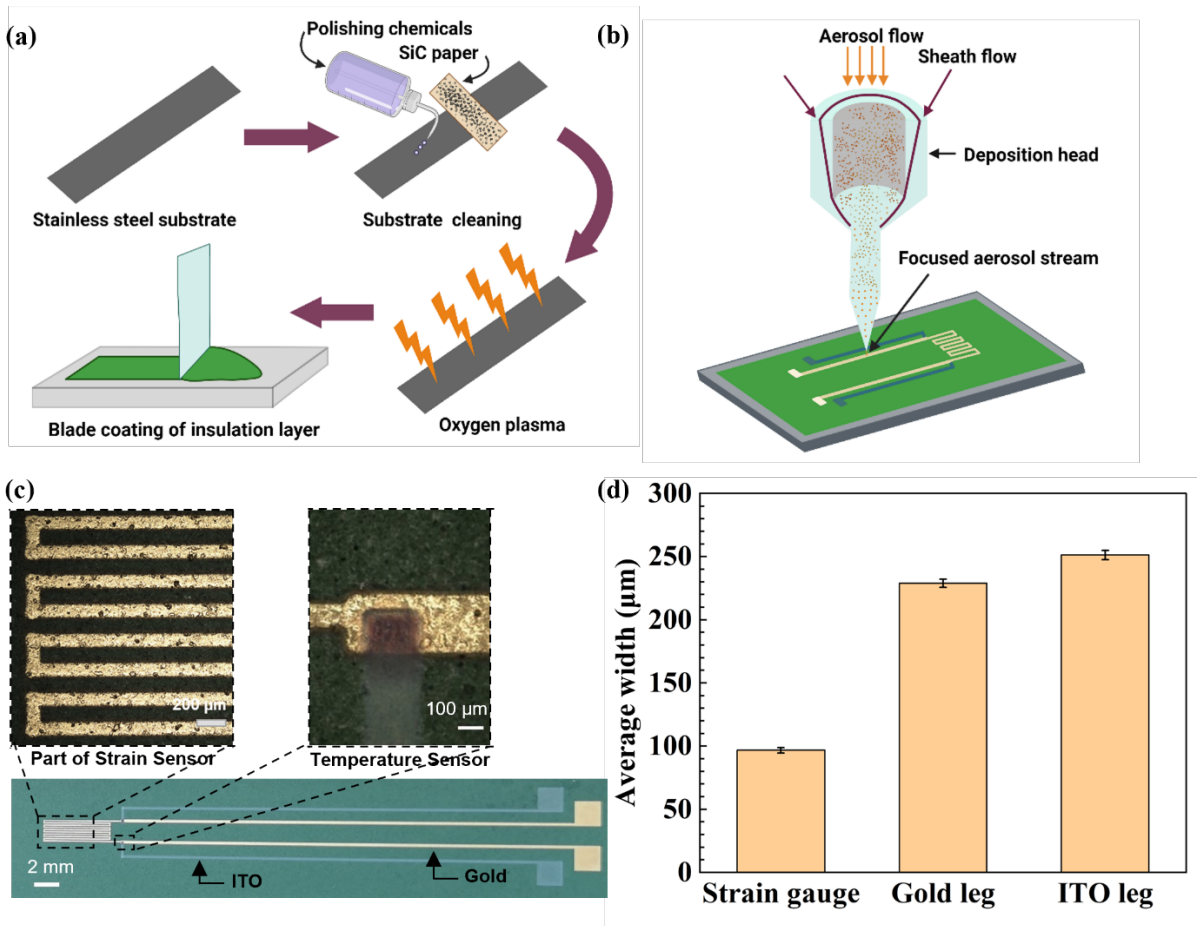

**Figure 1.** Schematic illustration of a) stainless steel substrate preparation and blade coating of a dielectric layer on the substrate, b) aerosol jet printing of multi-modal sensor on stainless steel substrate (not to scale), c) optical image of the printed sensor consisting of two thermocouples (gold ITO junction) and one strain gauge (gold) along with image viewed at a high magnification of temperature sensor and portion of strain sensor, d) width fluctuation of the printed sensor in distinct regions.

Substrate preparation and the AJP printing process for the multimodal sensor are depicted schematically in Figure 1. To prepare the stainless-steel substrate (154 mm × 26 mm × 0.5 mm) for printing, it is initially cleaned with isopropyl alcohol and acetone to remove any surface impurities. The surface is then abraded using 200-grit silicon carbide paper to eliminate the possibility of adhesion, such as flaking paint, corrosion, scale, etc. A wet abrasion process is employed using chemical polishing agents and 400-grit silicon carbide paper. Then, the substrate is subjected to a 5-minute oxygen plasma treatment at 100 W (PDC-001-HP, Harrick Plasma, Ithaca, NY) to remove any unwanted foreign particles from the substrate surface. A dielectric layer of HG1 ceramic cement with a thickness of 94.4 ± 4.8 µm (measured by DektakXT, Bruker profilometer, 0.2µm stylus radii, and 1 mg stylus force) is applied to the substrate by blade coating to ensure electrical insulation between the sensor and the substrate. The HG1 ceramic cement is then cured through four steps, including air drying for 30 minutes, thermal curing at 200 °C for 30 minutes, holding at 400 °C for 30 minutes, and finally, holding at 600 °C for 30 minutes.

The multimodal sensor is then printed on the cured substrate by an aerosol jet printer (Optomec, Inc., AJ 300 system, Albuquerque, NM). Prior to sensor printing, an AutoCAD file illustrating the sensor design shown in Figure S1 in SI is fed into the printer to create the toolpath for printing. The AJP system includes a deposition head, a programmable motion system, an ultrasonic atomizer, and a pneumatic atomizer. Fabrication of a multimodal strain sensor with integrated thermocouples necessitates using two materials with high-temperature stability and distinct Seebeck coefficients. This study employs gold and ITO nanoparticle inks. The first step is to load the ultrasonic atomizer with a solution containing 2 wt. % gold nanoparticles dissolved in xylene (UTD Au25X, gold nano ink in xylene, UT DotsInc. Champaign IL). The ultrasonic atomizer generates aerosolized ink droplets ranging from 0.5 to 5 microns[47, 48] to print the strain sensor. The aerosolized ink droplets are collimated into a micro-jet by an annular sheath gas ($N_2$) that surrounds the entering stream at the transition zone of the printhead. Once the strain sensor is printed using the gold ink, the ink vial of the ultrasonic atomizer is filled with a 5 wt. % ITO nanoparticle solution (particle size of 18 nm, water solvent, ≥99.5% trace metals basis, Sigma Aldrich Co.) to print the ITO electrode. The printed ITO and Gold electrodes form a thermocouple junction (Figure 1c) due to their distinctive Seebeck coefficient. The sensor is constructed by depositing three layers of ink and using multiple lateral passes to create individual electrodes. Real-time pictures of sensor printing at different printing steps are shown in Figure S2. The printing process parameters, such as sheath gas flow rate, carrier gas flow rate, printing speed, standoff distance, ultrasonic power for ink aerosolization and droplet formation, platen temperature are optimized for the gold and ITO inks (Table S1). Following printing, the printed sensors are sintered in a tube furnace at 800 °C for 2 hours in the air to consolidate the printed nanoparticles into a dense, thermally stable, electrically conductive, and mechanically robust structure.

**2.2. Calibration Methodology.** To evaluate the strain measured by the printed multimodal sensor, the stainless-steel substrate is fixed at one end using a bracket attached to the timing belt of a vertical shaft while the other end is kept free, resembling a cantilever beam (as shown in Figure S3). At first, the sensor is calibrated using a commercial strain sensor (Omega, Precision strain gauges, Resistance 350 Ω, Gauge factor 2.05) and commercial thermocouple (Omega, K-type). The commercial strain gauge is attached to the stainless-steel substrate at exactly the same position as the printed strain gauge and then flexed up to 10 mm in 5 steps using motorized linear movement of a timing belt. The resistance at each deflected position is measured. Strain at different bending positions can be determined from the relative resistance change with respect to resistance at the initial position and gauge factor of the commercial strain gauge using the following equation

$$GF = \frac{\Delta R/R_0}{\varepsilon} \qquad (1)$$

where ε represents strain, $\Delta R/R_0$ represents the change of resistance with reference to the initial resistance $R_0$, and GF is the gauge factor, which represents the sensitivity of the strain sensor. Cantilever beam theory is also applied to calculate the strain experienced by the cantilever beam at each bending position. The theoretical strain calculated from beam theory and strain measured by the commercial strain sensor are compared to ensure the accuracy of calibration methodology. Details of the theoretical calculation of strain is shown in Figure S14. The measured strain is then used to obtain the gauge factor of the printed multimodal sensor. The Seebeck voltage generated at the printed thermocouple junction is also meticulously calibrated using the commercial K-type thermocouple. The DC Seebeck voltage generated at the printed thermocouple was measured up to 550 °C and at the same time the hot and cold junction temperature was measured by the commercial thermocouples. Thus, the generated voltage was calibrated against the known temperature difference between the hot and cold junction.

**2.3. Simultaneous Temperature and Strain Measurement Setup.** To measure the resistance of the printed strain gauge, a high-frequency alternating current is applied through the printed gold electrodes to the strain gauge, and the resulting AC voltage is measured via the printed ITO electrodes to measure the AC resistance change and determine the strain. At the same time, the DC Seebeck voltage generated at the printed ITO and gold electrode junction is measured to determine the temperature difference between the ITO-gold thermocouple junction and the cold side of the thermocouple. The experimental setup is configured to simultaneously detect strain and temperature as illustrated in Figure S3-4 in the supporting information.

The customized test setup integrates an electric muffle furnace (KSL-1100X-S, 950W, Maximum temperature 1100°C, MTI Corporation) to perform measurements at high temperatures. The system comprises a stepper motor to drive the timing belt, a microcontroller driver to control the motor, and a base to hold the vertical shaft, to which the timing belt and stainless-steel bracket are attached. A 34970A data acquisition/data logger switch unit is used to measure the Seebeck voltage generated across the gold and ITO junction and the resistance of the strain sensor. The data acquisition unit is controlled by a computer, which is also used to monitor and record data. The sensor is clamped with the bracket, which can move up and down with the linear motion of the timing belt. The movement of the timing belt is precisely regulated at each testing cycle to ensure accurate testing results. The sensor is deflected like a cantilever beam as the tip is placed beneath a ceramic tube attached to the furnace wall, when it moves up strain is induced in it. A rectangular section (30 mm × 12 mm) is cut from the furnace door to insert the sensor inside the furnace and allow up to 10 mm deflection. The furnace door is closed after inserting the sensor inside the furnace and then high-temperature super wool thermal insulation paper is used to minimize heat loss from the cut-away section of the furnace door during high-temperature measurements. Moreover, a cooling water circuit is inserted at the stainless-steel bracket along with the sensor to keep the cold junction of the printed thermocouple at ambient temperature. To ensure thermal stabilization before conducting measurements at high temperatures, the multi-functional sensors undergo annealing at 600°C for 15 hours.

## 3. RESULTS AND DISCUSSION

### 3.1. Sensor Morphology.
Figure 1c shows an optical image of the sensor and a high-magnification image of the printed thermocouple junction and a part of the strain sensor. The printed sensor consists of two thermocouple junction and one strain gauge. Details of the multimodal sensor design, thermocouple hot junction, cold junction, etc. are shown in Figure S1. The width variation of the printed pattern at different regions of the sensor is measured under an optical microscope (63 locations in each region) and the corresponding average line width along with the standard deviation is shown in Figure 1d. Detailed line widths at different parts of the printed sensor are shown in Figure S5. The small standard deviation in the width variation indicates the high consistency of microscale additive manufacturing via aerosol jet printing. White light profilometer (Filmetrics, Profilm3D) is used to map the surface topology of the sensor and measure thicknesses at different parts as shown in Figure S6 and Figure S7, which reveals a mean thickness of 3.19 µm along the strain gauge and mean thicknesses of approximately 2.76 µm and 14.31 µm along the gold electrode and the ITO electrode respectively.

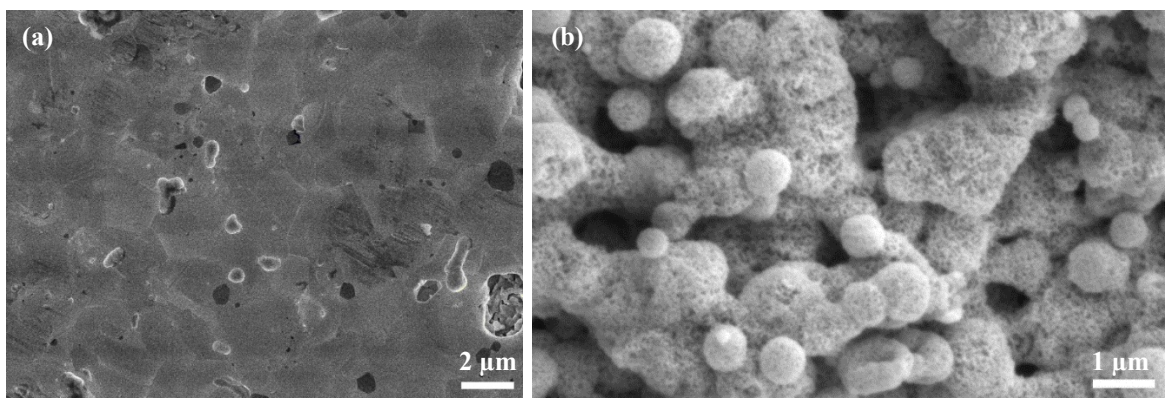

**Figure 2.** SEM images of the aerosol jet printed gold and ITO films sintered at 800 °C, a) morphology of the top surface of the gold film and b) morphology of the top surface of the ITO film.

Figure 2 shows the SEM images of the printed gold and ITO sintered at 800 °C. During sintering, the solvents are removed, resulting in porosity, which is then reduced by grain growth and particle coalescing to form well-connected and densified microstructures. The printed gold and ITO under optimized sintering conditions show electrical conductivity of $1.9 \times 10^7$ and $1.5 \times 10^4$ S/m, respectively. The gold film shows a highly dense microstructure with substantial grain growth because of the relatively high sintering temperature compared to the melting point of the gold nanoparticles, whereas the ITO shows a relatively large porosity as the sintering temperature is well below the melting temperature of ITO. The estimation of porosity is shown in Figure S8.

**3.2 Temperature Measurement.** The Seebeck voltages produced by the gold/ITO thermocouple junctions are measured at a wide range of temperature differences. The applied temperature gradient determines the thermoelectric voltage, which can be expressed as

$$V = \int_{T_c}^{T_h} S_G(T) - S_{ITO}(T) \, dT \tag{2}$$

Where $S_G$ and $S_{ITO}$ are the temperature-dependent Seebeck coefficients of gold and ITO electrodes, $T_H$ and $T_C$ represent the hot and cold junction temperatures, respectively. The hot side and cold side thermocouple junctions are shown in Figure S1. Two commercial thermocouples (Omega, K type) are inserted close to hot side and cold side of the sensor to continuously monitor and record the temperature. During measurement, the muffle furnace temperature is increased from ambient to 550 °C with a heating rate of 2.5 °C/min to increase the thermocouple hot junction temperature, and the corresponding Seebeck voltage is recorded. The Seebeck voltage increases linearly with the temperature difference between the hot and cold junctions of the printed thermocouple, as illustrated in Figure 3. The hot junction temperature, cold junction temperature, and generated thermoelectric voltage are shown in Figure S9. The cold junction temperature increases from 22 °C to 30 °C when the hot junction reaches 550 °C. The cold side is maintained as cold as possible using a liquid cold plate as shown in Figure S4. The printed thermocouples generated a maximum Seebeck voltage of 29.8 mV when the hot junction temperature reached 550 °C, with a temperature difference of 520 °C between the hot and cold junctions. The behavior of the printed thermocouple during heating and cooling indicates exactly the same ramp-up and down characteristics, as illustrated in Figure 3a.

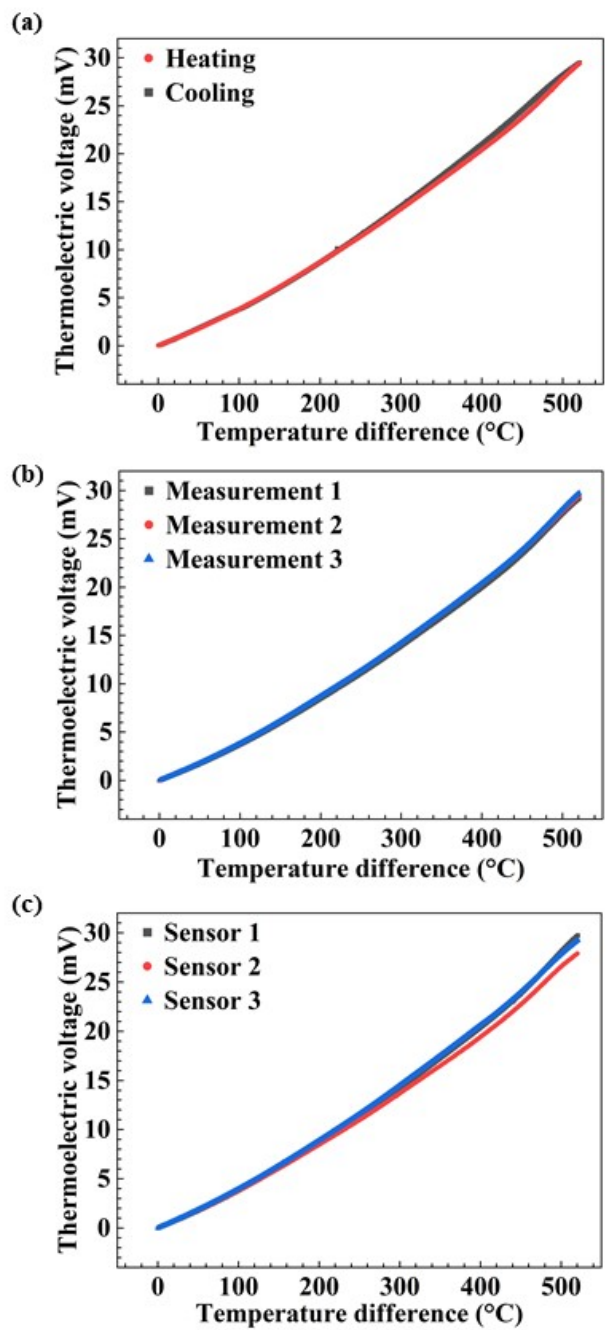

**Figure 3.** a) Thermoelectric voltage response of the printed gold/ITO thermocouple during heating and cooling cycles, b) stability of the printed thermocouple during three repeated measurements, and c) thermoelectric voltage output vs. temperature for three sensors printed under the same conditions.

For simplicity, only left thermocouple data has been shown in Figure 3 because the hot side temperature was measured near the left thermocouple. However, both the left and right thermocouple shows almost identical Seebeck voltage for same temperature difference as shown in Figure S10. The printed thermocouples show remarkable thermal stability and repeatability during repeated thermal cycling measurements at elevated temperatures up to 550 °C (Figure 3b), revealing high precision in temperature measurement. Moreover, the printed sensor demonstrates excellent reproducibility with three sensors showing an average thermopower of 55.64±1.5 µV/°C (Figure 3c). The printed thermocouple shows higher thermopower (sensitivity) than other thin film thermocouples reported in the literature (Table S2 in the supporting information).[2, 19, 49-52] The small variation of the thermopower between the three sensors can be attributed to the measurement error. The slightly variation in commercial thermocouple positions while testing different sensors can cause inaccuracy in exact hot and cold junction temperature measurement.

**3.3 Simultaneous Temperature and Strain Measurements at Room Temperature.** To obtain an accurate measurement of the strain gauge resistance without including lead wire resistances and contact resistance, the four-wire measurement method is used to measure the resistance of the strain gauge. An alternating current (AC) with a frequency of 5,000 Hz is input to the strain gauge via the gold electrode of the thermocouple, producing an AC resistive voltage that is measured across the ITO electrode. The resistance value of the strain gauge can be determined by the measured voltage and current of the AC signal. The DC Seebeck voltage across the thermocouple is also measured separately to determine temperature without any interference with the AC resistive voltage across the same thermocouple.

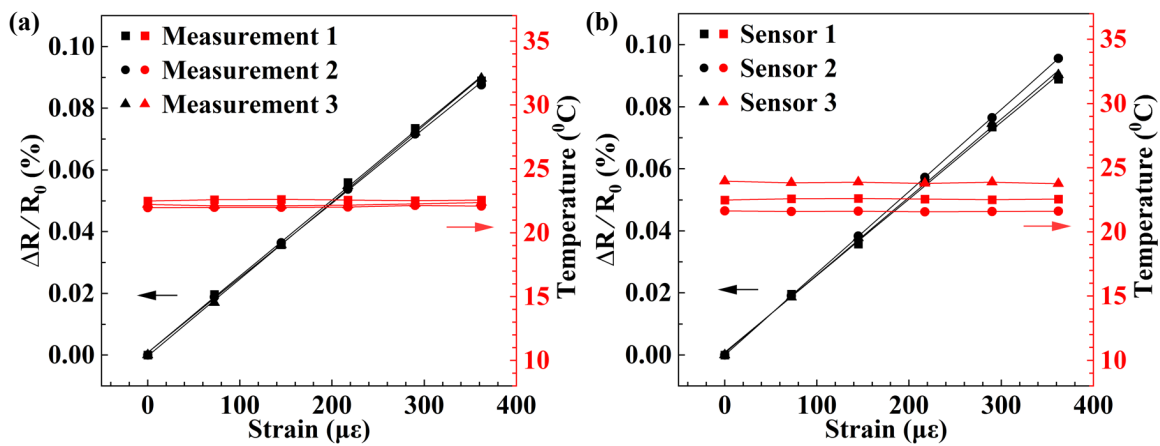

**Figure 4.** Simultaneous measurements of temperature and strain at ambient temperature, a) for three different measurements of a sensor to show the stability of the printed sensor, and b) for three different sensors.

Concurrent temperature and strain measurement results at room temperature of the printed multimodal sensors are depicted in Figure 4. The resistances of the three strain gauges are 11.95 Ω, 12.0 Ω, and 12.1 Ω at 22 °C ambient temperature. To obtain the response of the strain gauge at different strains, the sensor tip is deflected up to 10 mm in 5 equal steps in a cycle. The change of resistances with repeated bending cycles for a printed sensor is shown in Figure S11 in the SI. Figure 4a shows the relative change of resistance for three repeated measurements of the same sensor at different strains, along with the corresponding temperature, as measured by the multimodal sensor, demonstrating its high repeatability. Figure 4b shows high reproducibility with <3% variations for three different sensors printed under the same conditions. Moreover, the printed strain gauge has a gauge factor of 2.54±0.07 at room temperature, which is about 30% higher than that of most commercial strain sensors with a gauge value of around 2.[21, 22]

**3.4 Performance of the Multimodal Sensor at High Temperature.** Concurrent temperature and strain measurement results are illustrated in Figure 5a-d at temperatures 140 °C, 285 °C, 440

°C, and 540 °C respectively. The signals collected from the sensors were consistent and reliable across all temperature ranges in multiple measurements. The recorded thermoelectric voltage and a corresponding calibration curve of the sensor are utilized to determine the temperature measured by the printed thermocouple which agrees well with the commercial thermocouple.

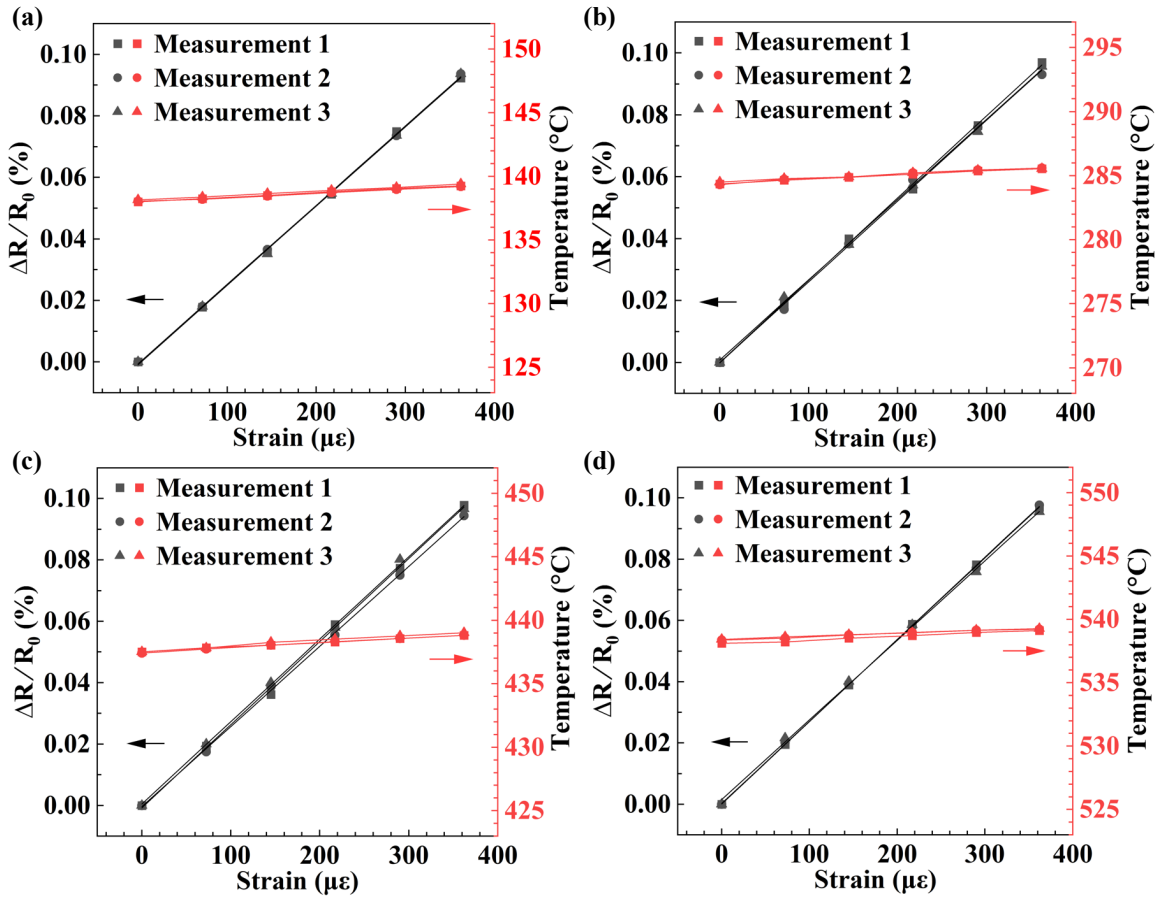

**Figure 5.** Simultaneous measurement of temperature and strain for three different measurements to show the stability of the multimodal sensor measured at a) 140 °C, b) 285 °C, c) 440 °C, and d) 540 °C.

To measure strain at a certain temperature, at first the temperature of the muffle furnace is stabilized for a long period of time to maintain the test strain gauge completely stable at that specific test environment, and then strain is induced to the strain gauge. Thus, the change of resistance of the gauge is ensured to be only due to the applied mechanical strain. Multiple measurements are conducted at different strains to assess the stability of the printed sensors. As shown in Figure 5, three black lines show the changes of resistances of same strain gauge in three different measurements and the red lines shows the three different temperature measurements. The repetitive results, as shown in Figure 5, demonstrate the high precision of the printed sensor that does not degrade even at 540 °C. The standard deviation of multiple data points at each strain is very low, indicating identical measurements of resistance changes for a given strain. As shown in Figure S12, the gauge factor of the printed strain gauge varies slightly with temperature. The variation of gauge factor with temperature is around 0.01%/K. The error bar associated with Figure S12 represents the standard deviation of gauge factor of three sensors measured at five different temperatures and indicates excellent repeatability of the three strain gauges, which ensures that the sensor signals can be accurately reproduced at high temperatures. Figure S13 shows that the resistance of the strain gauge increases from 12 Ω to 36 Ω with a temperature coefficient of resistance of 0.0038/°C when the temperature increases from ambient to 542 °C. This highlights the importance of in-situ temperature measurement to decouple the changes in resistance caused by temperature and strain when using resistive-based strain sensors.

## 4. CONCLUSION

In conclusion, gold and ITO nanoparticle inks were employed to fabricate a multimodal sensor for temperature and strain sensing simultaneously. The highly versatile aerosol jet printing process makes it possible to transform and integrate multiple nanoscale building blocks into a strain sensor with integrated thermocouples, enabling accurate strain measurement without complex temperature compensation methods. The printed multimodal sensor exhibits exceptional high-temperature stability up to 540 °C. The sensor demonstrates a competitive gauge factor of 2.54 at room temperature and a thermopower of 55.64 µV/°C over the gold and ITO thermocouple junction. The fully printed multimodal sensors demonstrated in this work open exciting opportunities to directly print and integrate emerging nanoscale materials into multifunctional devices for a broad range of applications.

## ASSOCIATED CONTENT

### Supporting Information

The Supporting Information is available free of charge on the ACS Publications website.

## AUTHOR INFORMATION

### Corresponding Author

*Email: yzhang45@nd.edu

### Author Contributions

The manuscript was written through contributions of all authors. All authors have given approval to the final version of the manuscript.

### Notes

The authors declare no competing financial interest.

## ACKNOWLEDGEMENTS

This work was supported by the U.S. Department of Energy under Award No. DE-NE0009138.

**Supporting Information**

# Aerosol Jet Printing of High-Temperature Multimodal Sensors for Strain and Temperature Sensing

*Md. Omarsany Bappy, Qiang Jiang, Stephanie Atampugre, and Yanliang Zhang\**

Department of Aerospace and Mechanical Engineering, University of Notre Dame, Notre Dame, Indiana 46556, USA

*Corresponding author: yzhang45@nd.edu

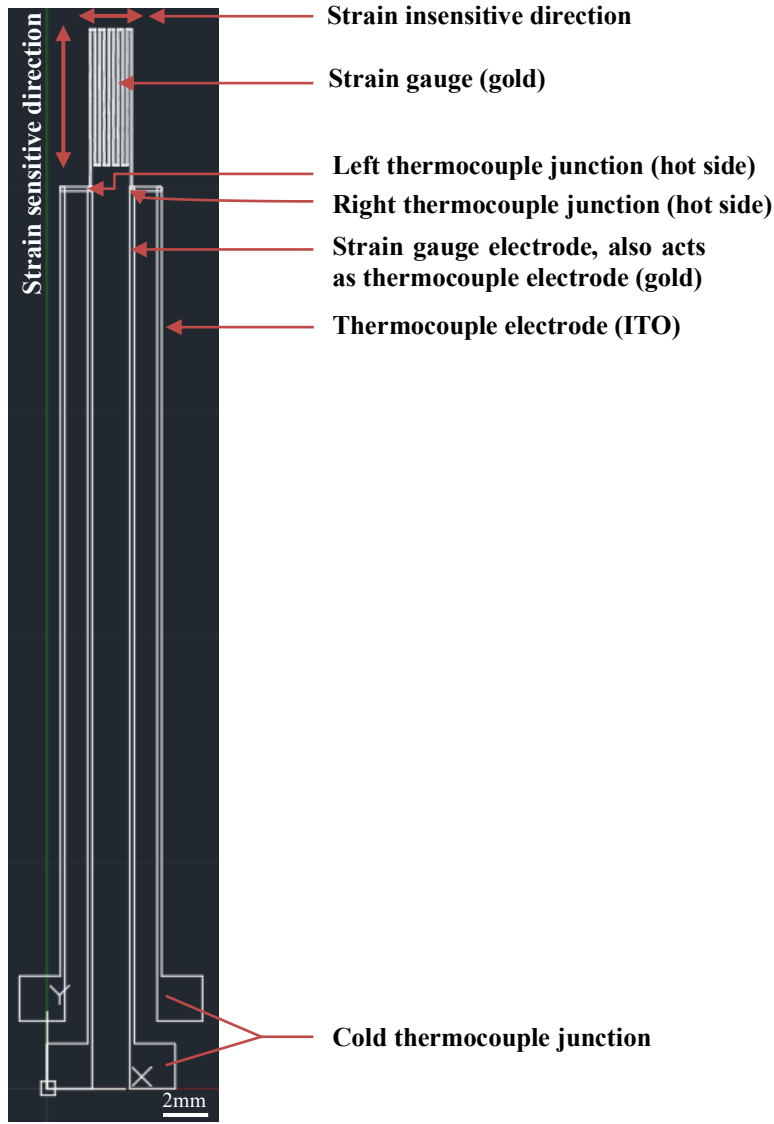

**Figure S1.** AutoCAD drawing of the printed multimodal sensor. This drawing is fed into the printer before printing the sensor.

The AutoCAD design of the multimodal sensor representing all the different parts along with the associated printed material has been mentioned. The two thermocouples have been named as left and right thermocouples. The hot side, as mentioned in the figure, is kept inside the oven during temperature measurement, and it acts as the hot junction of the thermocouple. On the other hand, the cold thermocouple junction, as mentioned in the figure, is kept outside the furnace and kept cold by cooling water flow during high-temperature measurement, as shown in Figure S4.

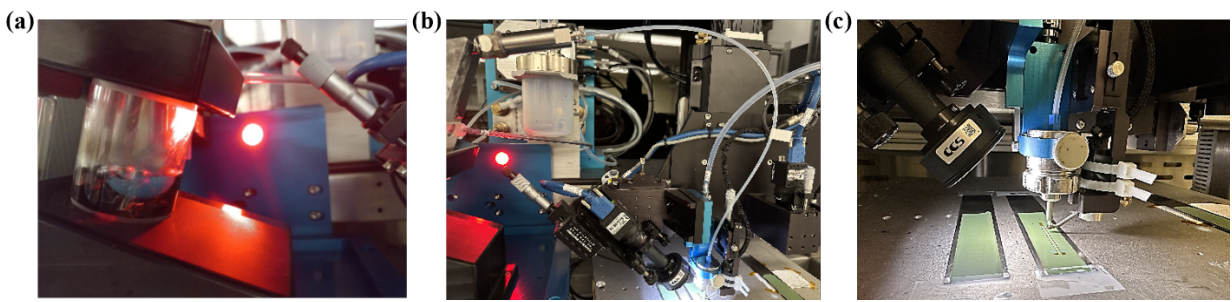

**Figure S2.** a) Atomization of Gold/ITO nano ink, b) transportation of aerosol droplet particles to the printer head, and c) ink deposition onto the substrate.

At first, 1.2 mL ink is taken in the ink vial and then set into the atomizer. To get the aerosol droplet from the nanoparticle ink, the ultrasonic atomizer current was set at 0.6 A for both gold and ITO. As soon as the atomizer power is turned on, the ink starts aerosolization. A real-time picture of the printing process of the bimodal sensor has been shown in Figure S2, where the atomization of Gold/ITO nano ink, transportation of aerosol droplet particles to the printer head, and finally, the deposition of ink onto the stainless-steel substrate have been captured. When the aerosol droplets start depositing through the nozzle tip, the printed line is taken onto a glass slide and seen under the microscope to check the line width, continuity, and overspray. Then, the printed parameters are optimized based on the perfect continuous line, desired line width, and minimal overspray. After optimization, the sensor AutoCAD file is fed into the printer to print the sensor onto the substrate.

**Table S1. Gold and ITO Nanoparticle Ink Printing Parameters**

| Parameter | Gold | ITO |
|---|---|---|
| **Nozzle diameter (μm)** | 200 | 200 |
| **Carrier gas flow rate (sccm)** | 20 | 26 |
| **Sheath gas flow rate (sccm)** | 100 | 60 |
| **Number of passes** | 3 | 3 |
| **Standoff distance (mm)** | 3 | 3 |
| **Platen temperature (°C)** | 85 | 85 |
| **Chiller temperature (°C)** | 16 | 16 |
| **Printing speed (mm/s)** | 2 | 2 |

The printing parameters can vary depending on the ink viscosity, particle size, solid content, ambient temperature and pressure, etc. That's why, before printing, it's always better to optimize the parameters.

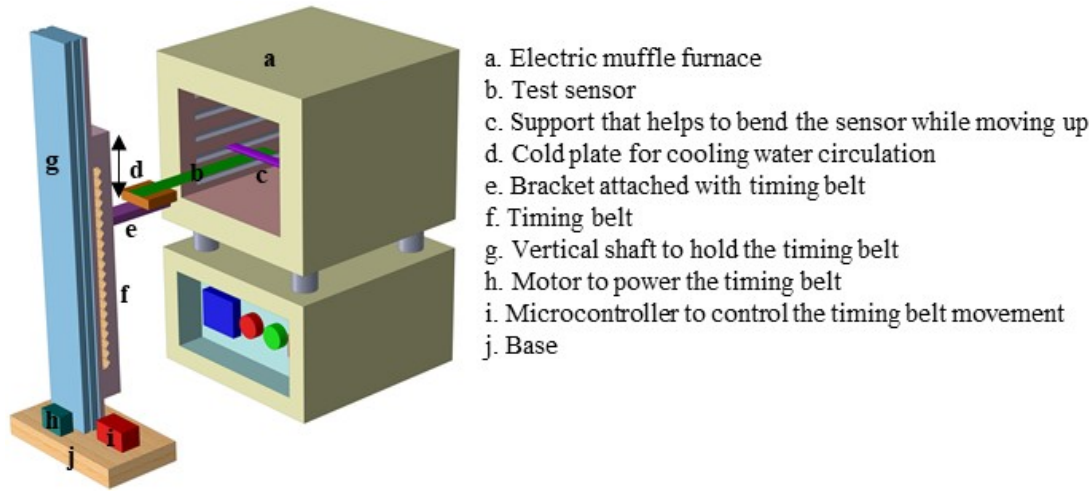

a. Electric muffle furnace
b. Test sensor
c. Support that helps to bend the sensor while moving up
d. Cold plate for cooling water circulation
e. Bracket attached with timing belt
f. Timing belt
g. Vertical shaft to hold the timing belt
h. Motor to power the timing belt
i. Microcontroller to control the timing belt movement
j. Base

**Figure S3.** Schematic of the experiment set-up for measuring concurrent temperature and strain at high temperature.

The schematic diagram represents the experimental setup used in this study to test the multimodal sensor. The sensor is fixed on top of the copper cold plate (denoted by 'd' in Figure S3) using a clamp to keep the thermocouple cold junction as cool as possible. Water is used as the coolant. The sensor tip is set underneath a ceramic tube (denoted by 'c' in Figure S3). When the timing belt moves up, strain is induced into the sensor (acts like a cantilever beam).

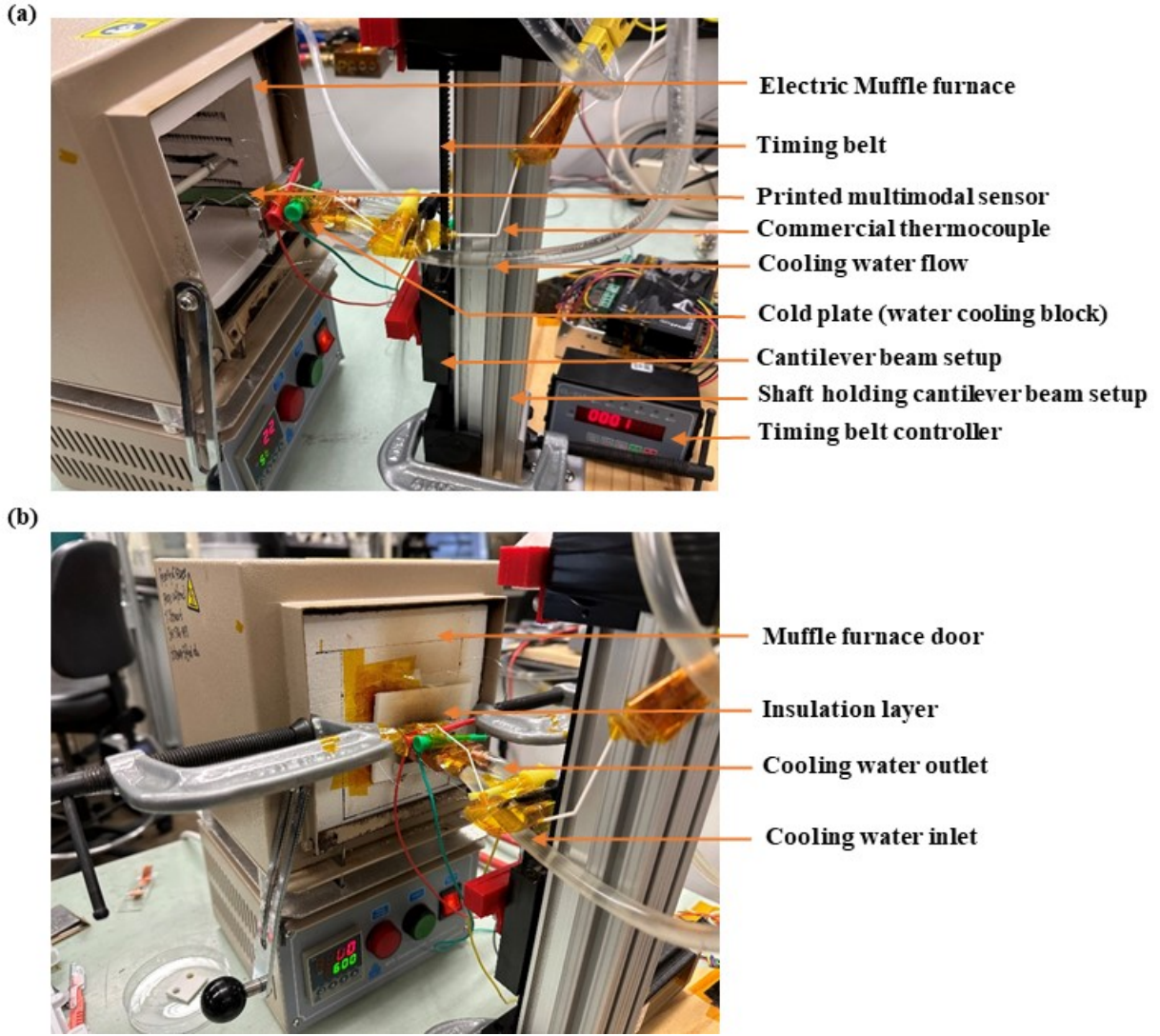

**Figure S4.** Real-time photo of the experimental setup during measurement at a) room temperature and b) high temperature.

Figure S4 (a) shows the measurement setup of strain testing at room temperature. To test the strain gauge, the timing belt is moved 10 mm upward first in 5 steps. Each step is set to take 3 minutes to get stable resistance data at that strained condition. Then the timing belt is moved 10 mm downward similarly to get the resistance value during straining down. During high-temperature measurements, the cooling water circuit is turned on to maintain the cold junction temperature as low as possible. The temperature at both the hot side and the cold side is continuously monitored and recorded.

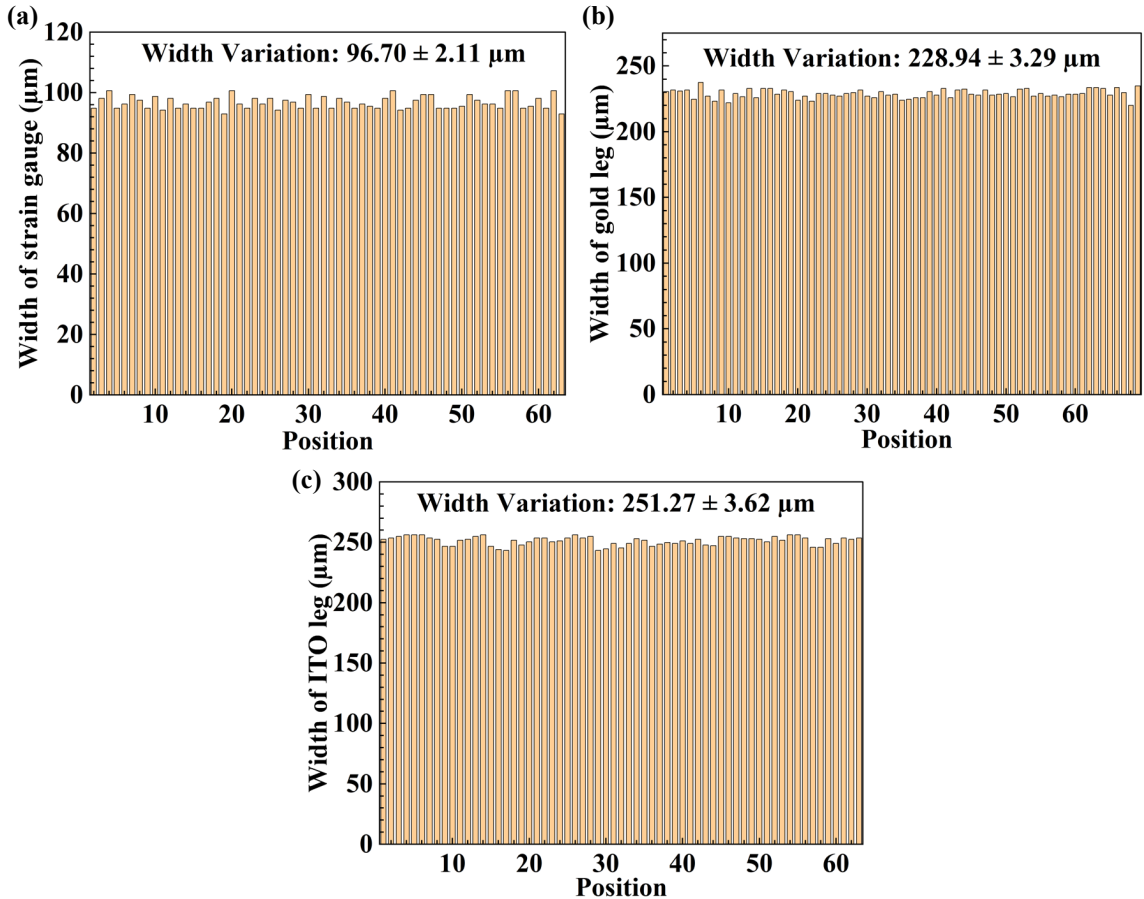

**Figure S5.** Width variation of the printed multimodal sensor at different regions a) variation of the width along the strain gauge is about ± 2.11 μm from the mean, i.e., 2.18% from the mean, b) variation of the width along the gold electrode is about ± 3.29 μm from the mean, i.e., 1.44% from the mean, and c) variation of the width along the ITO electrode is about ± 3.62 μm from the mean, i.e., 1.44% from the mean, indicating high resolution of the aerosol jet printed sensor.

After printing the sensors, the line width at different parts of the sensors was measured using an optical microscope. The low standard deviation of the printed lines shows the high-quality printing of functional devices by aerosol jet printers.

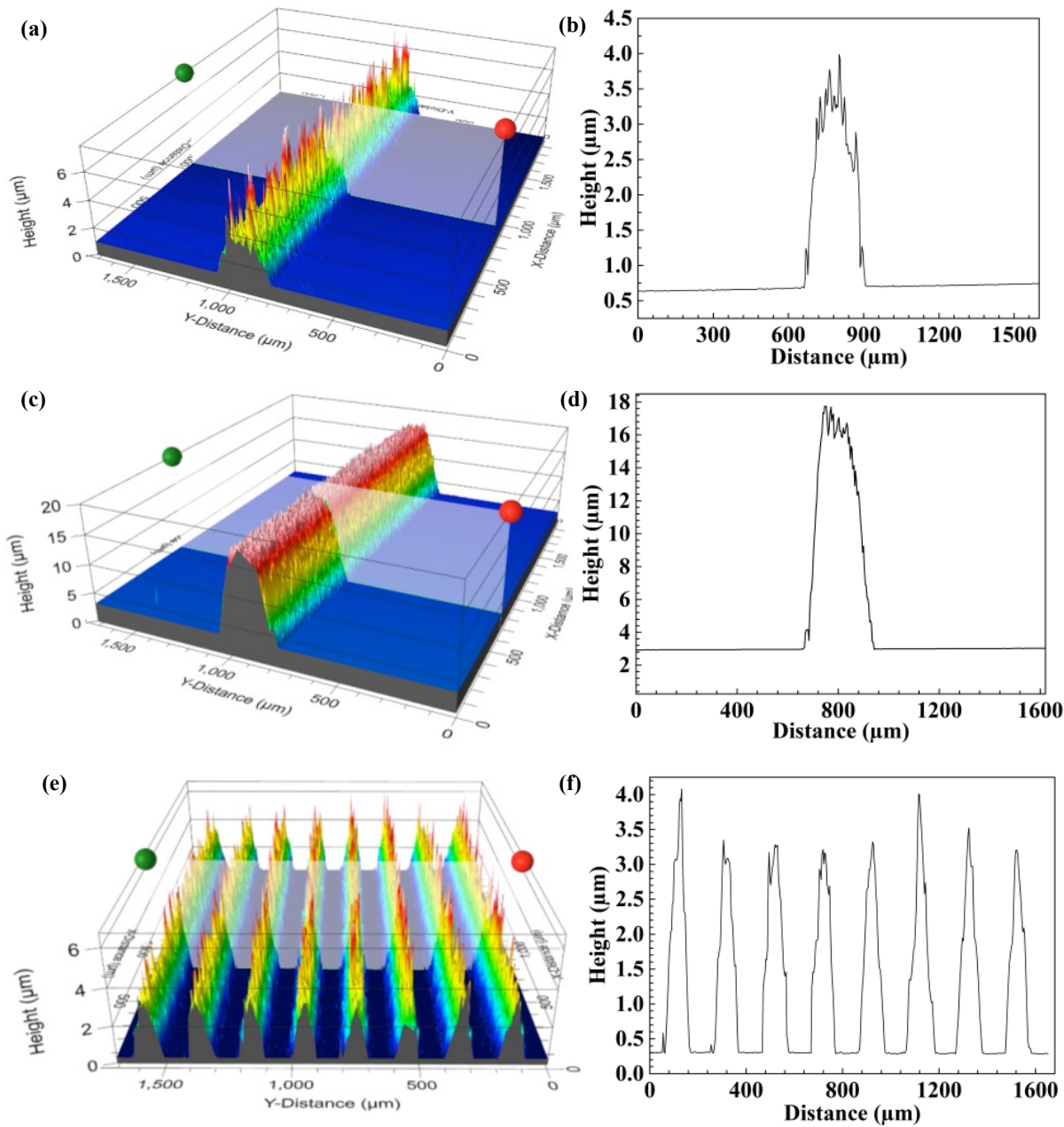

**Figure S6.** Thickness variation of the aerosol jet printed multimodal sensor at different locations using white light profilometer, a) a typical white light profilometer scan across the printed gold electrode, b) surface topology of the gold electrode, c) white light profilometer scan across the printed ITO electrode, d) surface topology of the ITO electrode, e) white light profilometer scan illuminating the surface topography of a strain sensor along the strain gauge, and f) thickness variation of the printed strain gauge at a certain plane.

Figure S6 shows the 3D scan of the sensor at different parts by using a White light profilometer (Filmetrics, Profilm3D). White light interferometry technology is used by this kind of profilometer to provide quantitative surface topological information. It is a nondestructive, non-contact type of measurement method offering good technology to measure film thickness. The film thickness shown in Figure S6 represents the film thickness across a plane only. Multiple film thicknesses are measured along different planes and then averaged to get the average thickness. Line width variation from bottom to top can also be observed by the profilometer image shown in Figure S6.

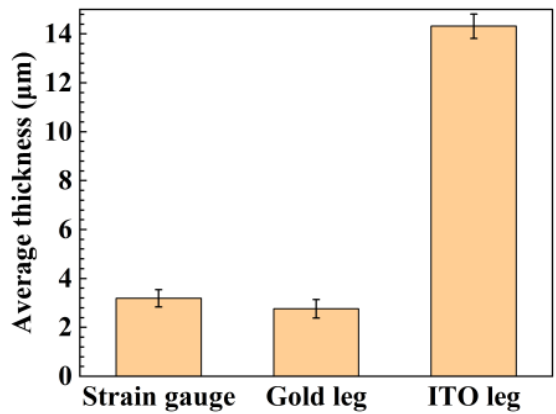

**Figure S7.** Thickness variation of the aerosol jet printed sensor along different parts.

The variation in film thickness along different parts of the printed multimodal sensor is shown in Figure S7. The average thickness shown in the figure is the average of the thickness taken in 5 different planes and then averaged, and the error bar represents the standard deviation of those 5 film thicknesses.

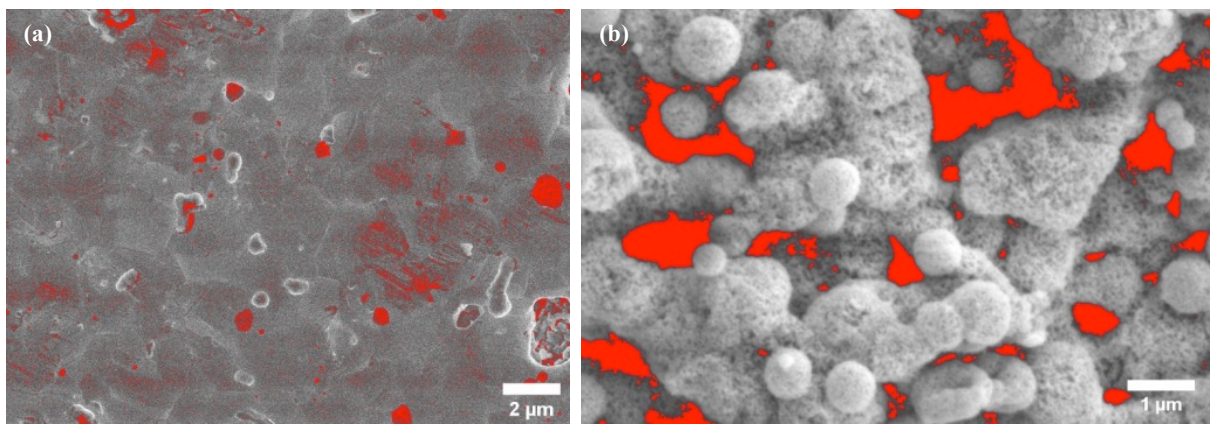

**Figure S8.** Estimation of porosity of aerosol jet printed (a) gold and (b) ITO sintered at 800°C.

The SEM image of the gold film shows that sintering close to the melting point of gold nanoparticles causes them to coalesce and form larger grains and grain boundaries and relatively dense structures with ~4% porosity. The SEM image of ITO shows a relatively large porosity of ~9% porosity as the sintering temperature is well below the melting temperature of ITO. The porosity has been estimated by ImageJ software.

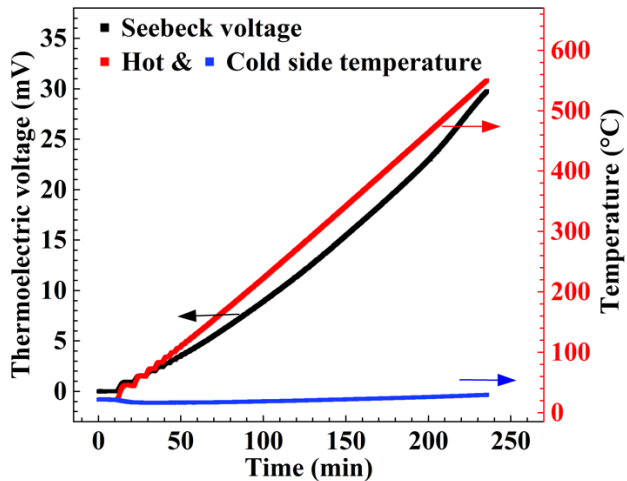

**Figure S9.** Hot side temperature, cold junction temperature, and corresponding thermoelectric voltage during a temperature measurement test have been shown.

Figure S9 shows the variation of thermoelectric voltage with temperature difference between the hot and cold thermocouple junctions. The hot and cold thermocouple junction temperatures, which were measured using a commercial K-type thermocouple, have been shown separately. Please note that thermoelectric voltage is a function of the temperature difference across the hot and cold junction.

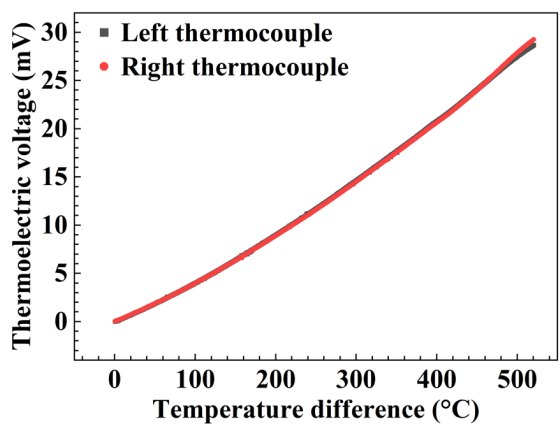

**Figure S10.** The thermoelectric voltage generated at the two thermocouples has been shown as a function of the temperature difference between the hot and cold junction of the printed sensor.

The Seebeck voltage generated at the left and right thermocouple of a sensor has been shown. During this measurement, the hot side temperature was measured near the left thermocouple. Both the two thermocouples of a single sensor show almost identical thermoelectric voltage across the same temperature difference.

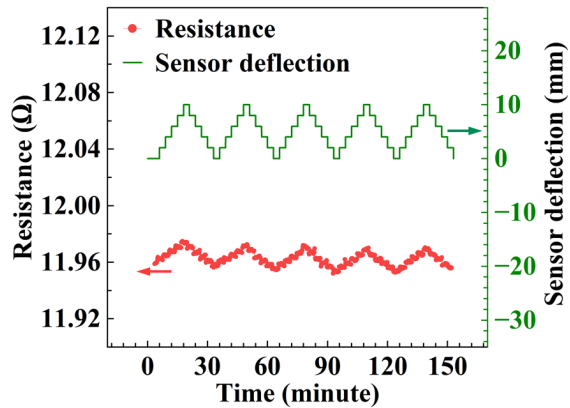

**Figure S11.** Variation of resistance of a strain gauge at different strains has been demonstrated during real-time monitoring. The sensor has been deflected up to 10 mm in 5 steps; both bending up and down have been performed, as shown by the green lines of the figure, and the corresponding change of resistance of the sensor has been shown.

The strain measurement results of a sensor at room temperature are shown in Figure S11. Initially, the sensor is at zero strain position, and the measurement is started. Then, after some time, the timing belt controller is turned on, and the sensor starts moving up or down according to the preset program. Strain is induced into the sensor as the sensor is bent, and its resistance changes. The resistance of the sensor at each bending position is shown in Figure S11.

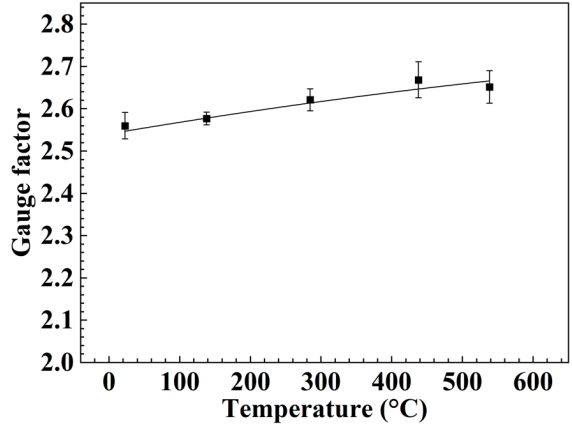

**Figure S12.** The effect of temperature on the gauge factor of the strain gauge during strain measurement shows a nonlinear fit, and the error bar represents the gauge factor variation at different temperatures.

The gauge factor varies (0.01%/K) with temperature, as shown in Figure S12. To measure the strain at different temperatures precisely, the gauge factor at that temperature should be known. We have tested the sensor at 5 temperatures, and the error bar in Figure S12 shows the standard deviation of the gauge factor of 3 sensors at that test temperature mentioned in Figure S12.

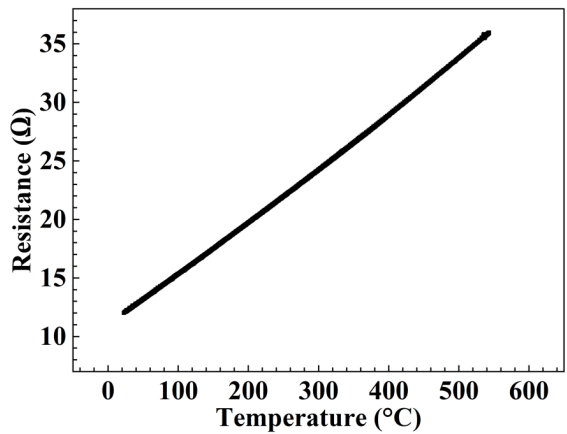

**Figure S13.** Variation of strain gauge (gold) resistance as a function of temperature. This graph helps to decouple the strain and temperature-sensitive resistance of the strain gauge.

As the temperature of the test sensor is increased from room temperature to 540 °C, the resistance of the printed gauge increases from 12 Ω to 30 Ω due to the metallic nature of gold.

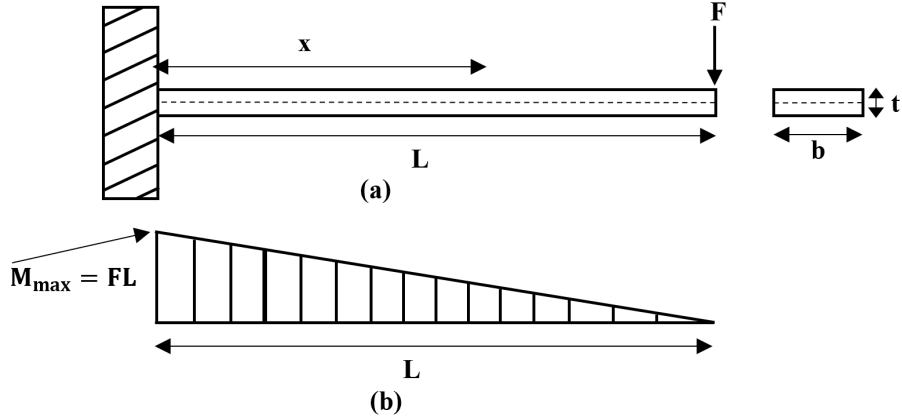

**Figure S14.** Strain calculation in a cantilever beam. (a) Side view of a cantilever beam. The left side is fixed, and the right side is free. (b) Moment distribution in the cantilever beam.

As mentioned earlier, when the sensor is deflected to induce strain, it acts as a cantilever beam. Here we will deduce the strain equation from Figure S14.

When load 'F' is applied at the beam tip, let the beam deflection be δ.

We can write, $\delta = \dfrac{FL^3}{3EI}$, where, E is Young's Modulus and I is moment of inertia.

Let's consider the strain gauge position at 'x' from the fixed end. So, the moment at any position of the beam will be $M = F(L - x)$ [considering clockwise moment as positive]

Now the bending stress at the beam is,

$\sigma = \dfrac{Mc}{I}$, where 'c' is the distance from the neutral axix $\bigl(\text{dotted line in Figure S14 (a)}\bigr)$

$$\therefore \sigma = \dfrac{F(L-x)c}{I} \quad \text{so,} \quad \dfrac{F}{I} = \dfrac{\sigma}{(L-x)c}$$

Now, $\delta = \dfrac{\sigma}{(L-x)c} \times \dfrac{L^3}{3E} = \dfrac{\epsilon \times L^3}{3(L-x)c}$, where strain, $\epsilon = \dfrac{\sigma}{E}$

$$\therefore \epsilon = \dfrac{3\delta(L-x)c}{L^3}$$

We print the sensor on top of the surface, so $c = t/2$ for our case. This is the theoretical equation of strain.

**Table S2. Comparison of the sensitivity of printed thermocouples manufactured by different techniques**

| Type | Fabrication Technique | Materials | Temperature Range (°C) | Sensitivity (μV/°C) | Ref. |
|---|---|---|---|---|---|
| **Thermocouple** | Aerosol jet printing | Gold/ITO | 20-600 | 55.64 | This work |
| **Thermocouple** | Screen printing | Silver/carbon black | 25-150 | 6 | 1 |
| **Thermocouple** | Lithography | Palladium/Chromium | ~150 | 21 | 2 |
| **Thermocouple** | Aerosol jet printing | Ti3C2Tx MXene/graphene | 20-200 | 53.6 | 3 |
| **Thermocouple** | Aerosol jet printing | Cu/Cu-Ni (50:50) | 30-232 | ~43 | 4 |
| **Thermocouple** | Screen printing | Carbon-Black inks | 50 ΔT | 1.1-5.3 | 5 |
| **Thermocouple** | Screen printing | In2O3/ITO | 1270 | 44.5 | 6 |
| **Thermocouple** | Screen printing | MoSi2/WSi2 | 1000 | 25.67 | 7 |